\documentclass[10pt]{forma}

\usepackage{graphicx} 
\def\be{\begin{equation}}
\def\ee{\end{equation}}

\def\bea{\begin{eqnarray}}
\def\eea{\end{eqnarray}}

\def\eg{{\it e.g.}}

\begin{document}
\hspace{1.0cm} 

\parbox{15.0cm}{

\baselineskip = 15pt

\noindent {\bf SIMULATION OF THE GRAVITATIONAL COLLAPSE AND 
FRAGMENTATION OF ROTATING MOLECULAR CLOUDS}

\bigskip
\bigskip

\noindent {\bf P. Berczik$^{1,2}$, M.I. Petrov$^1$}

\bigskip
\bigskip

\baselineskip = 9.5pt

$^1$\noindent {\small {\it Main Astronomical Observatory, National Academy of Sciences of Ukraine \\
27 Akademika Zabolotnoho Str., Kiev, 03680, Ukraine}} \\
\noindent {\small {\it e-mail:}} {\tt berczik@mao.kiev.ua; petrov@mao.kiev.ua}

\medskip

$^2$\noindent {\small {\it Astrophysics Group, Department of Physics, Rochester Institute of Technology \\
54 Lomb Memorial Drive, Rochester, NY 14623, USA}} \\
\noindent {\small {\it e-mail:}} {\tt berczik@cis.rit.edu}

\baselineskip = 9.5pt \medskip

\medskip 
\hrule 
\medskip

\noindent In this paper we study the process of the subsequent 
(runaway) fragmentation of the rotating isothermal Giant 
Molecular Cloud (GMC) complex. Our own developed Smoothed 
Particle Hydrodynamics (SPH) gas-dynamical model successfully 
reproduce the observed Cloud Mass-distribution 
Function (CMF) in our Galaxy (even the differences between the 
inner and outer parts of our Galaxy). The steady state CMF 
is established during the collapse within a free-fall timescale 
of the GMC. We show that one of the key parameters, 
which defines the observed slope of the present day CMF, is the  
initial ratio of the rotational (turbulent) and gravitational 
energy inside the fragmented GMC.

\medskip 
\hrule 
\medskip}

\vspace{1.0cm}

\renewcommand{\thefootnote}{\ }

\footnotetext{\copyright~P. Berczik, M.I. Petrov, 2004}

\renewcommand{\thefootnote}{\arabic{footnote}}

\baselineskip = 11.2pt

\noindent {\small {\bf INTRODUCTION}}
\medskip

\noindent SPH based 3D hydrodynamical codes, starting with the 
series of pioneering works by Monaghan and Lattanzio 
\cite{GM1983, LMPS1985, LH1988, ML1991}, are always very successfully 
applied to the study of evolution and fragmentation in molecular clouds 
and molecular cloud complexes. These early simulations have been 
usually performed with a few hundred to a few thousand of 
SPH particles and with a fixed (few parsec) spatial resolution. 

Nowadays the most up to date simulations of molecular cloud evolution 
(\eg~\cite{ML2000}) are performed using a few tens of thousands of SPH 
particles with variable smoothing lengths. These simulations also include 
the details of cooling and heating in the complex gas mixtures  
of H, H$_{2}$, CO and HII species.

Our present high resolution (64,000 SPH particle) simulations, with 
highly flexible and adaptive smoothing lengths, study the runaway collapse 
and the subsequent isothermal fragmentation of the isolated GMC complex with 
different rotational (turbulent) energy parameters of the clouds. The resulting 
CMF is compared with the recent observational distributions 
($d{\rm N}/d{\rm M} \sim {\rm M}^{-\gamma}$, where the slope of the 
power law $\gamma$ is in the range 1.4 to 1.8) of the 
molecular cloud complexes derived from the different CO data for the 
different parts of our Galaxy \cite{SSS1985, SRBY1987, KSRC1998, MT2000, DHT2001, HCS2001}. 


\bigskip
\noindent {\small {\bf METHOD}}
\medskip

\noindent Continuous hydrodynamic fields in SPH are described by the
interpolation functions constructed from the known values of these
functions at randomly positioned $N$ ``smooth'' particles with
individual masses $m_i$ \cite{M1992}. To achieve the same level
of accuracy for all points in the fluid it is necessary to use a
spatially variable smoothing length. In this case each particle
has an individual value of the smoothing length - $h_i$.

A more detailed and complete description of the basic numerical
equations of SPH can be found in many of our previous
publications (\eg~\cite{Ber1999, Ber2000, BHTS2002, BHTS2003, SBHTAFJ2004} 
and the references herein). Therefore we just briefly 
repeat the skeleton SPH equations of the code here. The density at the 
position of the particle $i$ can be defined as:

$$
\rho_i~=~\sum_{j=1}^N m_j~\cdot~W_{ij},
$$

%
%
%

\newpage

\begin{figure}[htbp]
\centerline{%
\begin{tabular}{c@{\hspace{0.25in}}c}
\includegraphics[width=3.2in]{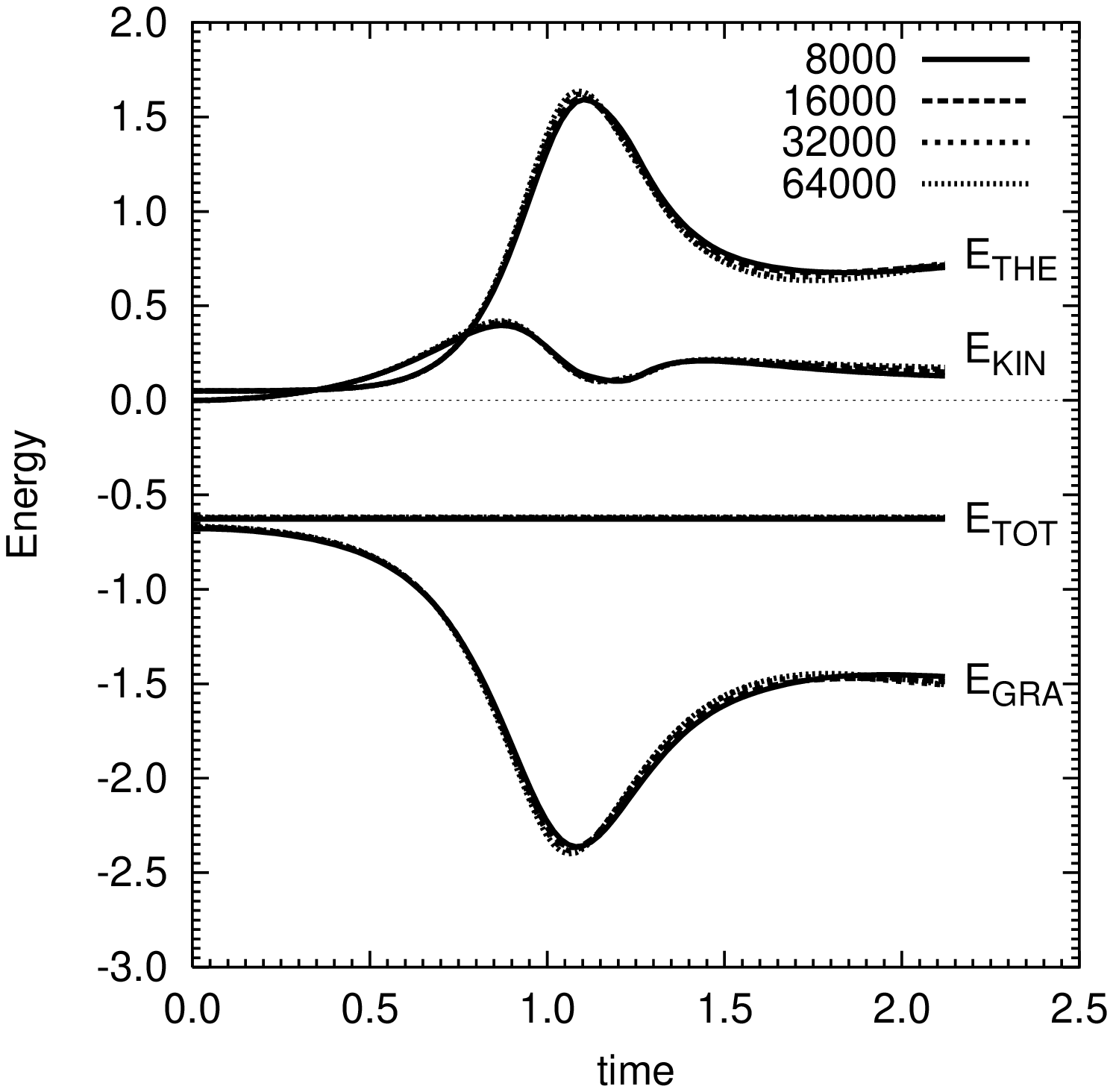} & \includegraphics[width=3.2in]{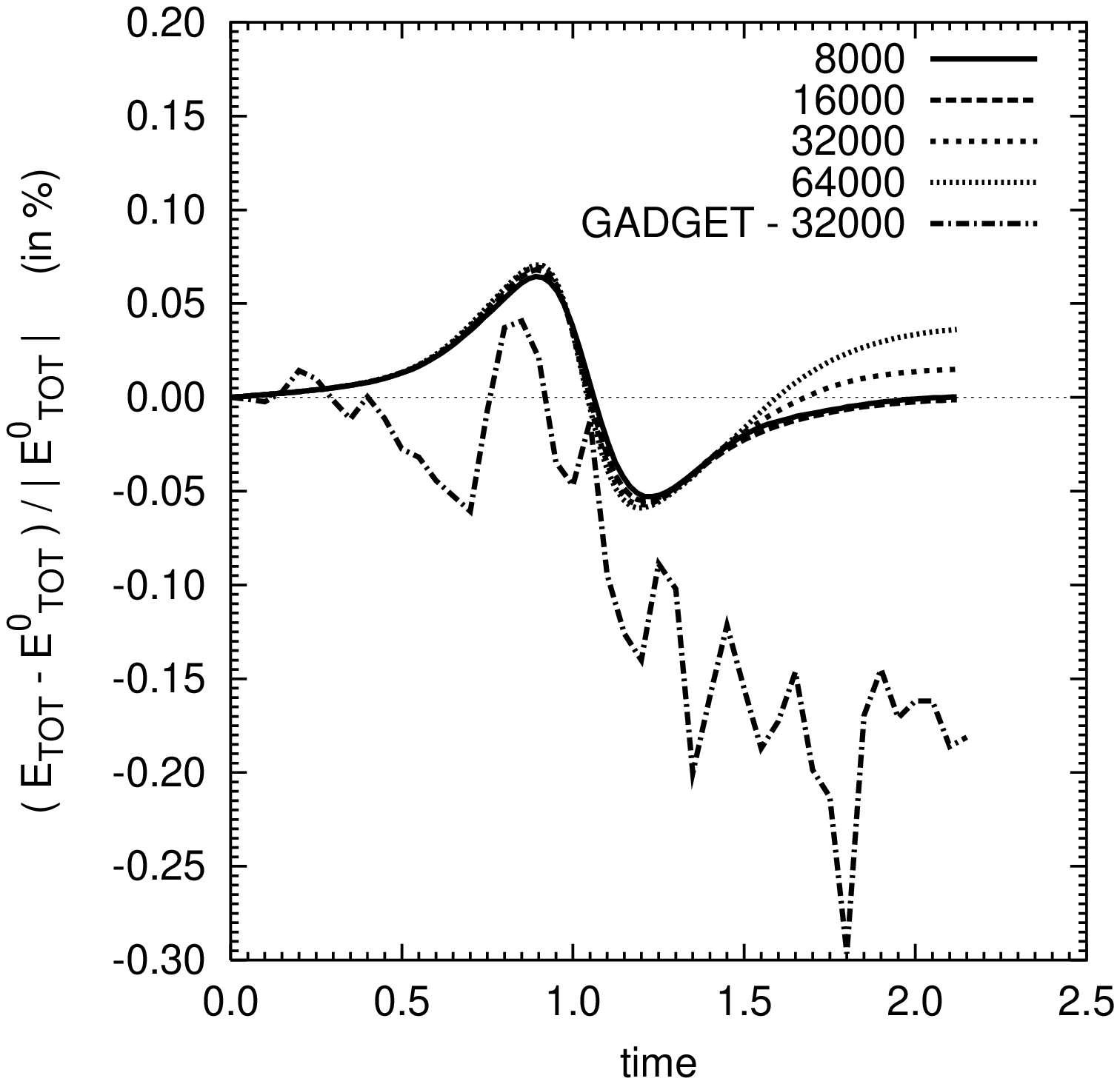} \\
a.~Evolution of the energy. & b.~Relative total energy error.
\end{tabular}}
\caption{Time evolution of the thermal, kinetic, potential and
total energy for the adiabatic collapse of an initially ``cold'' gas
sphere. The different lines corresponds to the different
gas particle numbers. We also present the
energy error results of the {\tt GADGET} code with
``standard'' parameters for 32,000 particles.}
\label{contr}
\end{figure}

The equations of motion for a particle $i$:

$$
\frac{d \bf{r}_{\it i}}{d t}~=~{\bf v}_i, 
$$

$$
\frac{d \bf{v}_{\it i}}{d t}~=~-~\sum_{j~=~1}^{N} m_j \left(\frac{P_i}{\rho_i^2}~+~\frac{P_j}{\rho_j^2}~+~\tilde{\Pi}_{ij}\right)~\nabla_i~W_{ij}~-\nabla_i~\Phi_i.
$$

\noindent where $P_i$ is the pressure, $\Phi_i$ is the self gravitational 
potential and $\tilde{\Pi}_{ij}$ is an artificial viscosity term. 

%
%
%
%
%

The internal energy equation has the form:

$$
\frac{d u_i}{d t}~=~\frac{1}{2}~\sum_{j~=~1}^{N} m_j~\left( \frac{P_i}{\rho_i^2} + \frac{P_j}{\rho_j^2}~+~\tilde{\Pi}_{ij} \right)~ ({\bf v}_i - {\bf v}_j)~\nabla_i ~W_{ij}~+~\frac{\Gamma_i~-~\Lambda_i}{\rho_i}.
$$

Here $u_i$ is the specific internal energy of the particle
$i$. The term $(\Gamma_i - \Lambda_i)/\rho_i$ accounts for non
adiabatic processes not associated with the artificial viscosity.
We present the radiative cooling in the form proposed by
\cite{NMN2000} (see case ``B'') using the {\tt MAPPINGS III}
software \cite{SD1993}:

$$
\Lambda~=~\Lambda(\rho,~u,~{\rm Z},~\ldots)~\simeq~\Lambda^{*}(T,~{\rm [Fe/H]})~\cdot~n_i^2, ~~~~~~~~~~n_i~=~\rho_i/(\mu~\cdot~m_p),
$$

\noindent where $n_i$ is the hydrogen number density, $T_i$ the
temperature and $\mu$ the molecular weight. 


The equation of state must be added to close the system:

$$
P_i~=~(\gamma - 1)~\rho_i~\cdot~u_i, 
$$

\noindent where $\gamma$ is the adiabatic index.


In SPH one of the basic tasks is to find the nearest neighbors of
each SPH particle, i.e. to construct an interaction list for each 
particles. Basically we need to find all particles with $\mid
{\bf r}_{ij} \mid \leq 2 \max(h_i,~h_j)$ in order to estimate the
density and also calculate the hydrodynamical forces.

In our code we keep the number of neighbors exactly
constant by defining $2 h_i$ to be the distance to the $N_B$ -
nearest particle. The value of $N_B$ is chosen
such that a certain fraction of the total number of ``gas''
particles $N$ affects the local flow characteristics. From these
we need to {\tt SELECT} the closest $N_B$ particles. Fast
algorithms for doing this exist \cite{PTVF1995}. 
For computational reasons, if the defined $h_i$ becomes smaller
than the selected minimal smoothing length $h_{min}$, we set the
value $h_i~=~h_{min}$.

To calculate the self gravitational potential $\Phi_i$ and self
gravitational force $-\nabla_i \Phi_i$ we use the 
Mitaka Underground Vineyard (MUV) {\tt GRAPE6} computer system at 
the National Astronomical Observatory of Japan [{\tt http://www.cc.nao.ac.jp/muv/}].
For a more detailed description of the {\tt GRAPE6} board and fot links to 
publication about the {\tt GRAPE6}, we refer the reader to the official 
homepage of Jun Makino at Tokyo University
[{\tt http://grape.astron.s.u-tokyo.ac.jp/$\sim$makino/grape6.html}].

For the time integration of the system of hydrodynamical equations
we use the second order Runge-Kutta-Fehlberg scheme. The time
step $\Delta t_i$ for each particle depends on the particle's
acceleration ${\bf a}_i$ and velocity ${\bf v}_i$, as well as
on the sound speed $c_i$ and the heating {\it vs.} cooling balance:

$$
\Delta t~=~C_n~\cdot~\min_{i} \left[~\sqrt{\frac{2 h_i}{\mid {\bf a}_i \mid}};~~
\frac{h_i}{\mid {\bf v}_i \mid};~~
\frac{h_i}{c_i};~~
\frac{u_i}{\dot{u}_i}~\right],
$$

\noindent where $C_n$ is the Courant's number $=~0.1$. For 
computational reasons we fix the minimal integration time
step $\Delta t_{min}$. 

%
%
%
%
%
%

The main aim of our current work is a detailed study of the isothermal fragmentation processes inside the collapsing ``cold'' molecular cloud complex. For the purpose of finding the fragments and its physical parameters (mass and size) we use our own cluster finding algorithms. In our algorithms we modify the well known and ``standard`` friend-of-friend (FOF) method \cite{HG1982}. Instead of just using the particle positions in the process of ``constructing'' or finding the clusters (fragments) we also use the information about the density distribution inside each potential cluster. On the basis of the density distribution analysis we can finally select in the more accurate way the members of our fragments (clusters). 
In this sense our method is more close to the so called {\tt SKID} method, which is well described at the homepage of the Washington University ``N-body Shop'' [{\tt http://www-hpcc.astro.washington.edu/tools/skid.html}]. Here the reader can find a more detailed description of this density base method which is specially designed to find the gravitationally bound groups of particles in the N-body like simulations. 

One of the features of our cluster finding routine is in the setting of the minimum limit of gaseous particles to 5, in order for a fragment to form. In other words we don't count as ``real'' a cluster where the number of particles is less then 5. 

\begin{figure}[t]
\centerline{%
\begin{tabular}{c@{\hspace{0.25in}}c}
\includegraphics[width=3.2in]{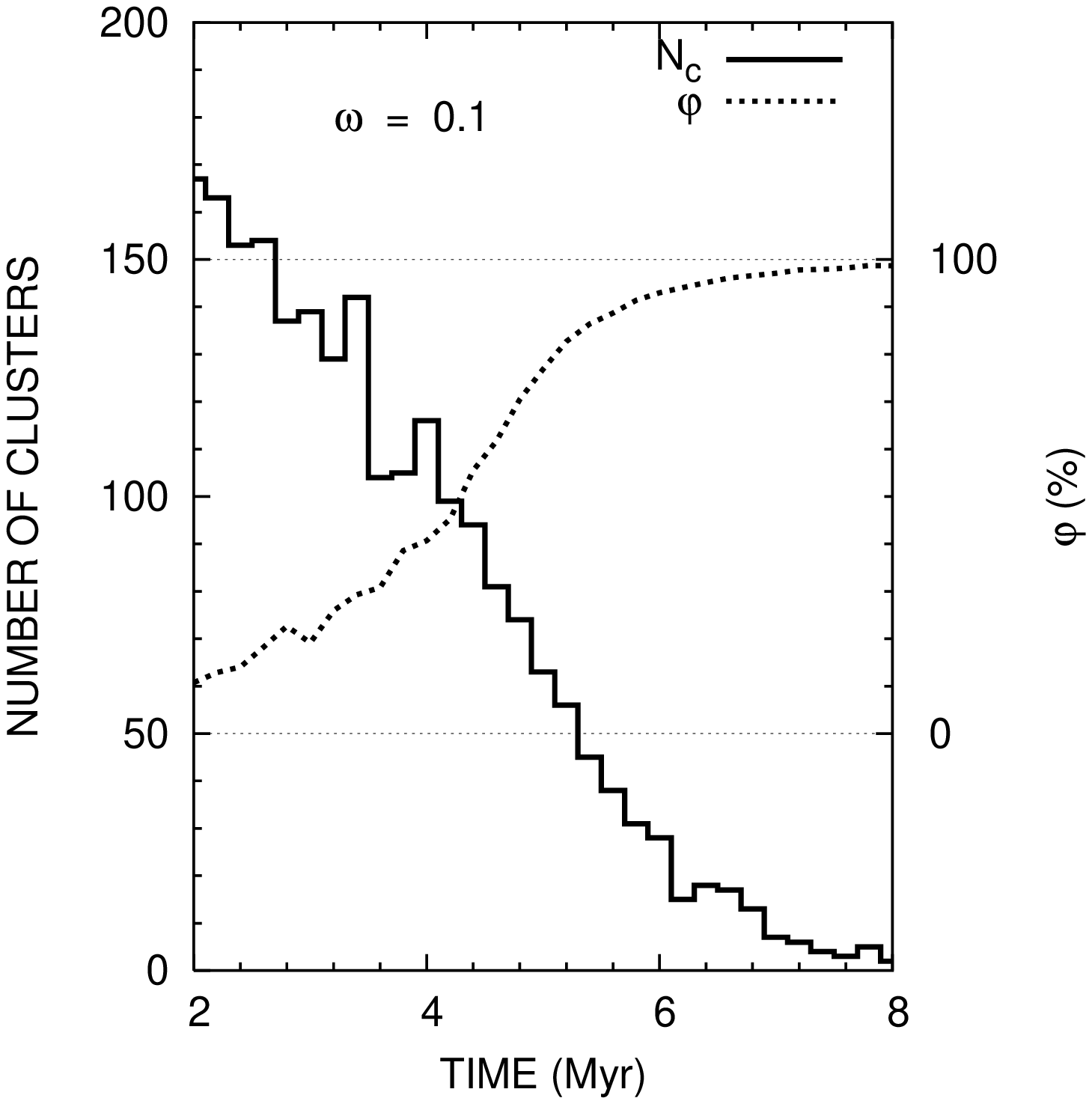} & \includegraphics[width=3.2in]{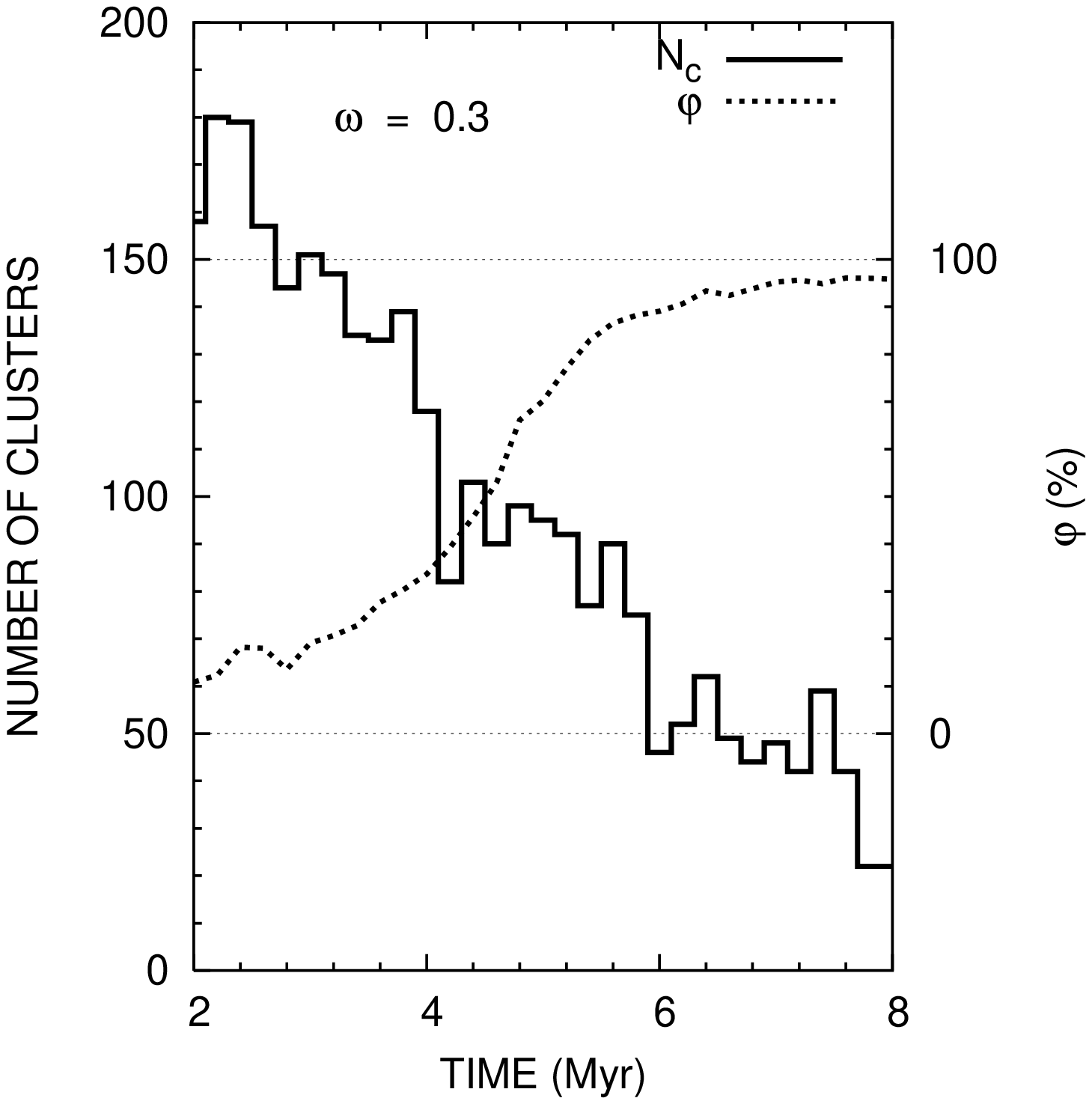} \\
a.~``Slow'' initial rotation $\omega = 0.1$. & b.~``Fast'' initial rotation $\omega = 0.3$.
\end{tabular}}
\caption{The time evolution of the total number of clusters N$_{\rm c}$ and the total mass fraction inside these clusters $\phi$ during the simulations. Starting from the $\sim$ 5 Myrs more than 80 \% of the total mass is already concentrated inside the fragments. At around 6 Myrs already almost 95 \% of the total mass is inside the clusters.}
\label{cluster}
\end{figure}


\bigskip
\noindent {\small {\bf CODE TESTING}}
\medskip

\noindent The self­gravitating collapse of an initially isothermal ``cold'' gas
sphere has been a common test problem for different SPH codes
\cite{E1988, HK1989, SM1993, CLC1998, TTPCT2000, SYW2001}.
Following these authors, for the testing of our code, we calculate the 
adiabatic evolution of the spherically symmetric gas cloud of total mass M and radius R. 
For the initial internal energy per unit mass we set the value: 
$u~=~0.05~\frac{{\rm G}~{\rm M}}{{\rm R}}$. The initial density profile of the 
cloud calculates as: 
$$
\rho(r)~=~\frac{{\rm M}}{2~\pi~{\rm R}^2}~\frac{1}{r}.
$$

\begin{figure}[t]
\centerline{%
\begin{tabular}{c@{\hspace{0.25in}}c}
\includegraphics[width=2.8in]{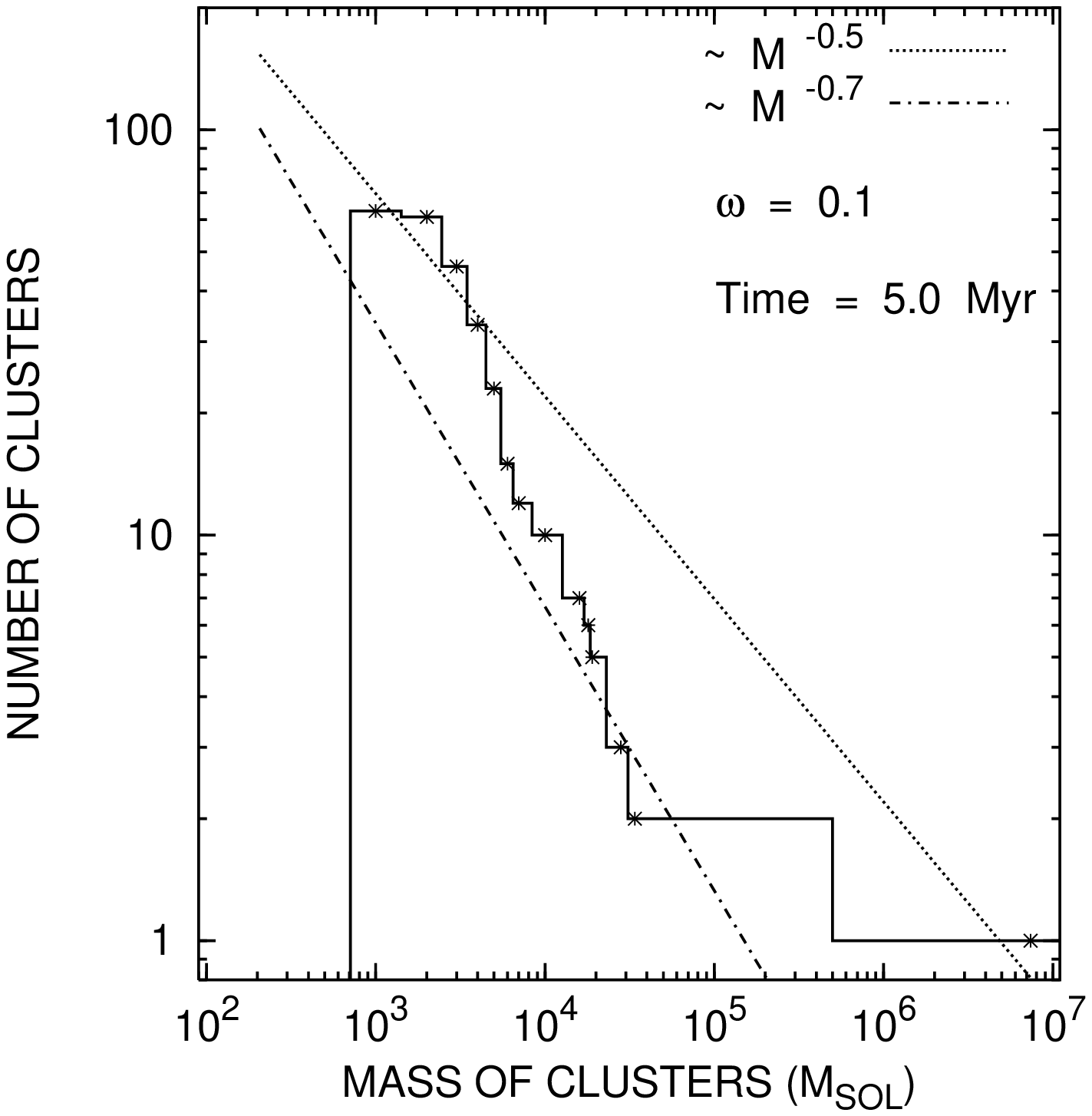} & \includegraphics[width=2.8in]{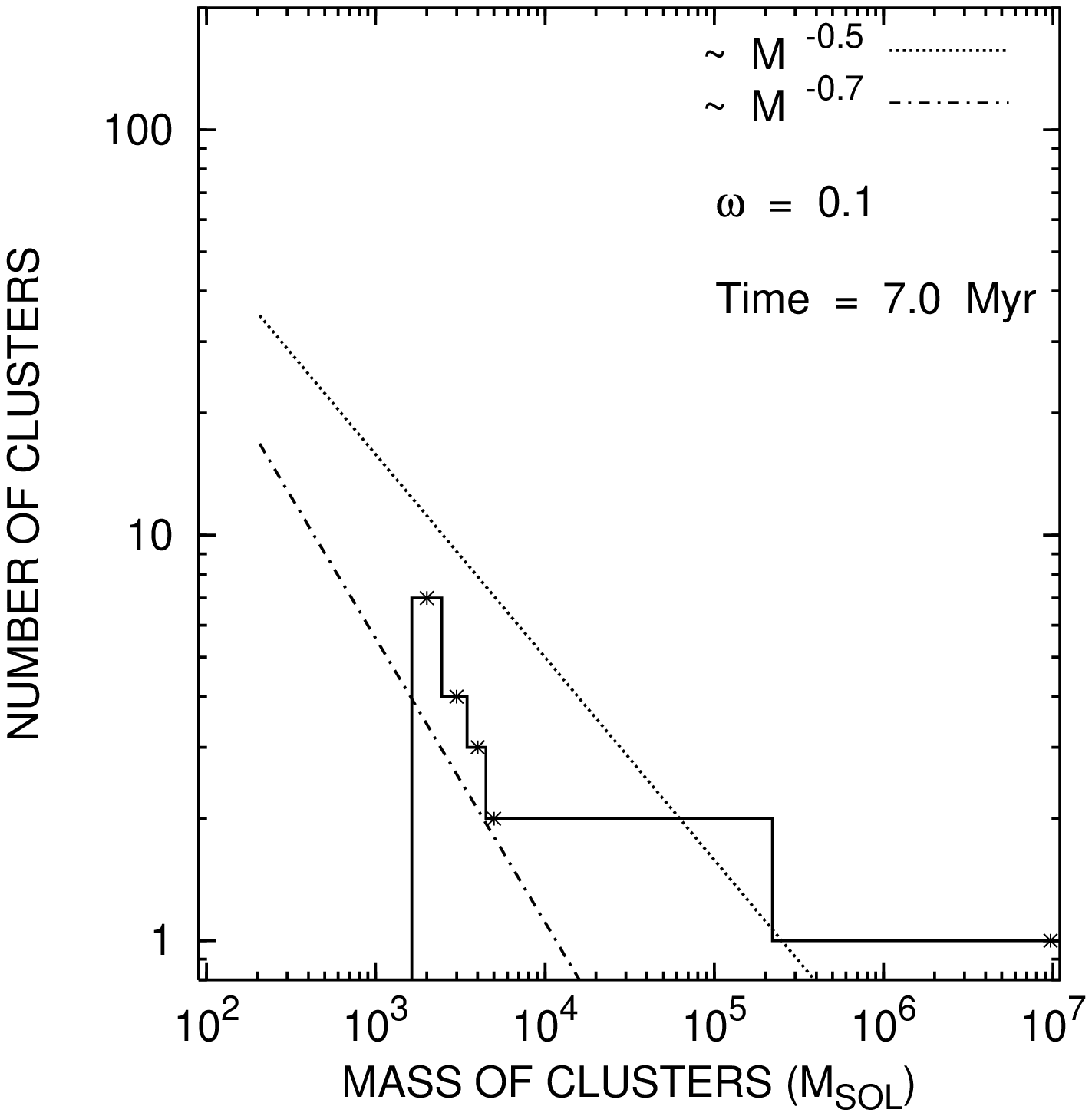} \\
a.~``Slow'' initial rotation $\omega = 0.1$, Time = 5 Myr. & b.~``Slow'' initial rotation $\omega = 0.1$, Time = 7 Myr. \\
\includegraphics[width=2.8in]{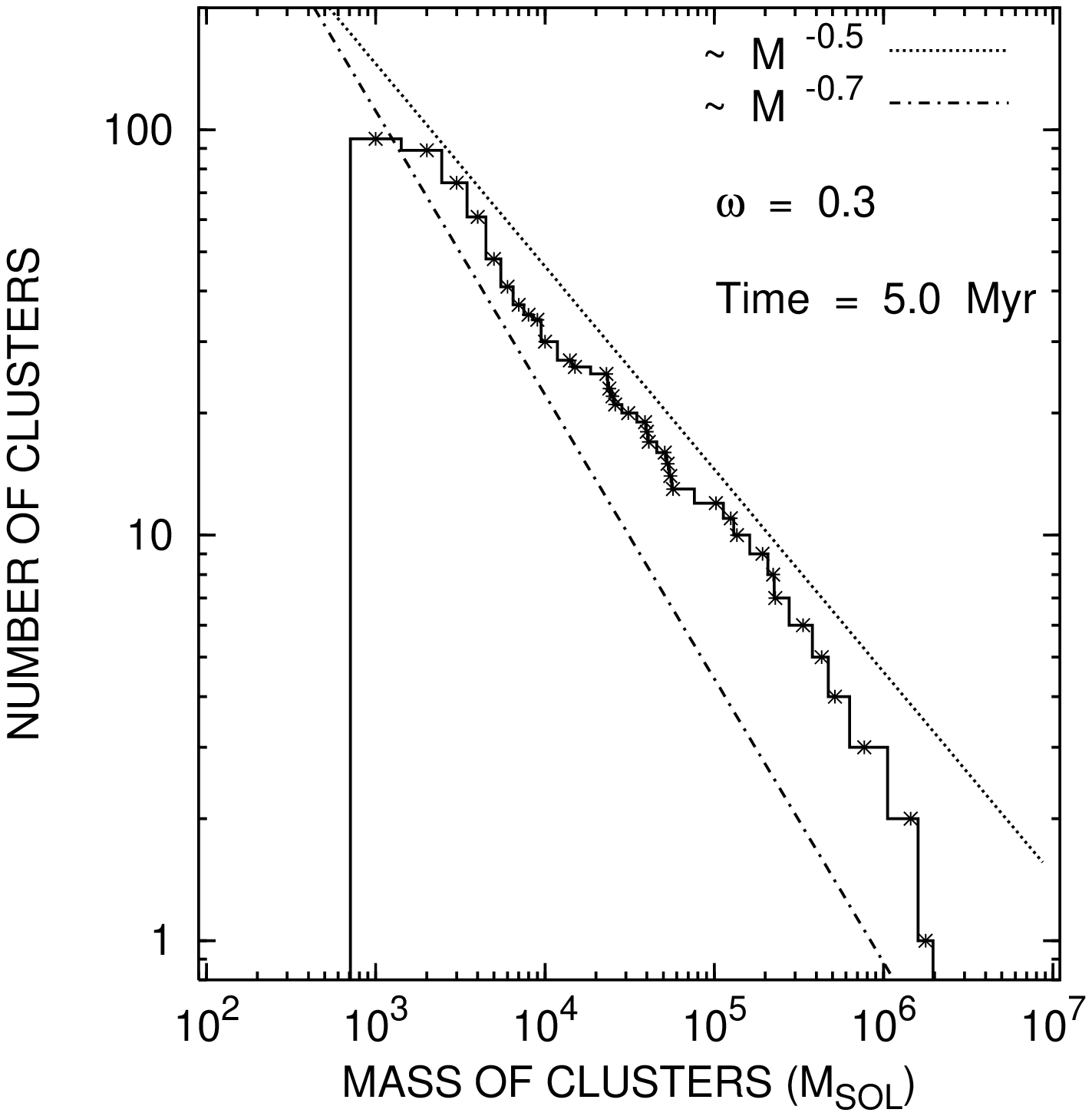} & \includegraphics[width=2.8in]{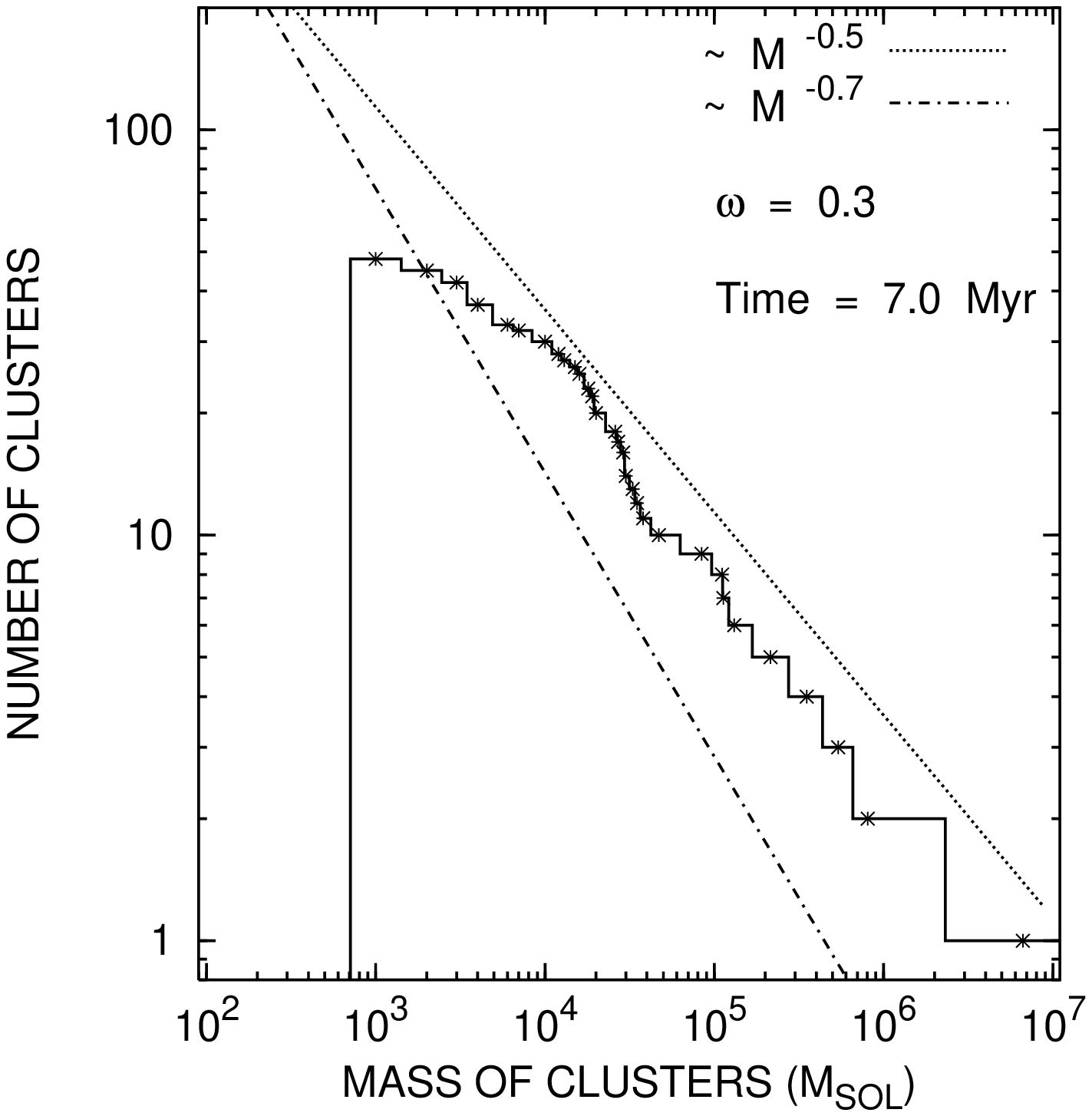} \\
c.~``Fast'' initial rotation $\omega = 0.3$, Time = 5 Myr. & d.~``Fast'' initial rotation $\omega = 0.3$, Time = 7 Myr.
\end{tabular}}
\caption{Four snapshots of the integrated cluster distribution function (ICMF).}
\label{icmf}
\end{figure}

We distribute randomly the gas particles inside the set of
spherical shells in a manner that reproduces the initial density 
profile. At the start of the simulation the gas particles are at rest. 
For the presentation of the results we use a system of units 
where G~=~M~=~R~=~1.

In Fig.~\ref{contr} we show the time
evolution of the different types of energy and the relative total
energy error during the calculation. For comparison of our 
test results we also plot the energy error 
results from the serial variant of the {\tt GADGET} public access 
SPH-TREE code \cite{SYW2001} with ``standard'' 
parameters for the 32,000 particles 
[{\tt http://www.mpa-garching.mpg.de/gadget/}]. 

During the central bounce around $t \approx 1.1$ most of the
kinetic energy is converted into heat, and a strong shock wave
travels outward. For all of these runs the number of neighbors 
was set $N_B~=~50$ and the gravitational softening was set 
$\varepsilon~=~0.01$. For the integration of the system of 
equations we use the second order Runge-Kutta-Fehlberg scheme 
with a fixed time step $\Delta t~=~10^{-4}$.

The results presented in Fig.~\ref{contr} agree very well with those
of \cite{SM1993} and \cite{SYW2001}. The maximum relative
total energy error is around 0.05~\% even for moderate (8,000)
particle numbers. The largest adiabatic test calculation (with 64,000 gas
particles up to $t \approx 2.2$) on an Intel Pentium 4 (3.4 GHz) 
host machine with a {\tt GRAPE6} board took $\approx$ 3.67 days of 
total CPU time.


\begin{figure}[t]
\centering
  \includegraphics[width=5.5in]{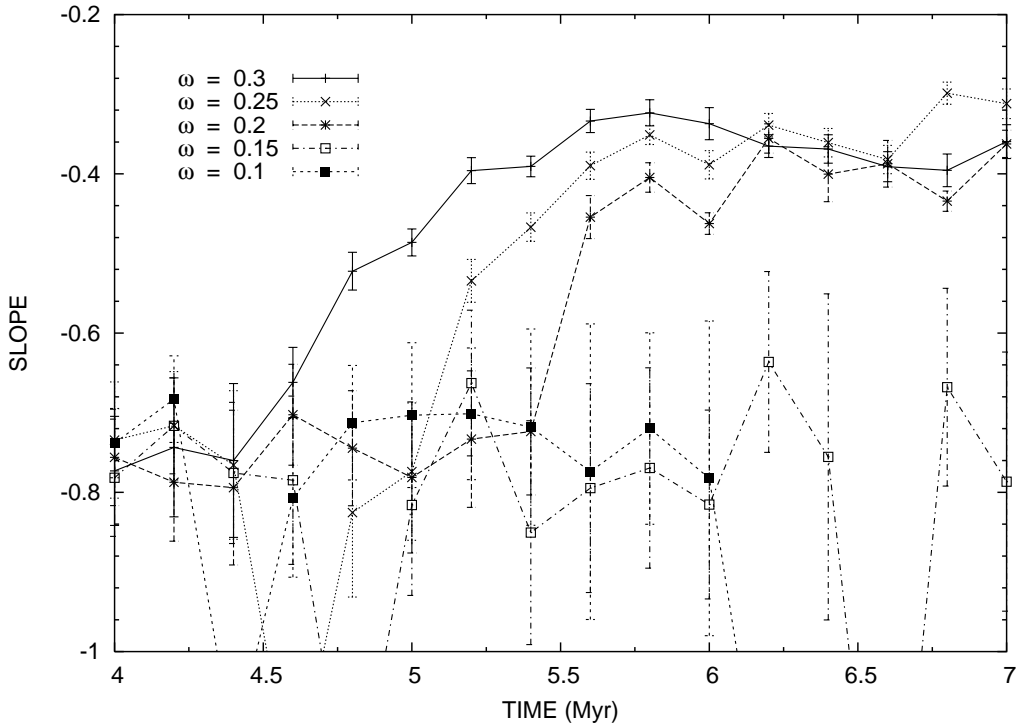}
  \caption{The ICMF slope time evolution for our models with different rotation parameters.}
  \label{slope}
\end{figure}


\bigskip
\noindent {\small {\bf INITIAL CONDITIONS}}
\medskip

\noindent As an initial condition for our molecular cloud fragmentation study we 
use a model in which the parameters are comparable with the largest GMC complexes 
in our Galaxy \cite{MT2000, DHT2001, HCS2001}. For the mass of the system we set 
M$_{cloud}$ = 10$^7$~M$_{\odot}$. For the radius of the cloud we set 
R$_{cloud}$ = 100 pc. For an initial density distribution we use the previous 
formula where $\rho(r)$~$\sim$~$\frac{1}{r}$. For the purpose of checking the 
possible ``resolution'' effects we carry out two sets of runs with 32,000 
and 64,000 gas particles (with the corresponding indexes ``low'' and ``high''). 
The total gravitational energy of the system in such a case can be easily 
calculated using the simple formula: 

$$
{\rm E}^0_{\rm GRA} = -\frac{2}{3}~{\rm G}~{\rm M}^2_{cloud}/{\rm R}_{cloud}.
$$

For the initial temperature we set the value which produced the overall 
ratio of the thermal energy to the gravitational energy of the system 
at the fixed level $\alpha$ $\equiv$ E$^0_{\rm THE}$/$\mid$E$^0_{\rm GRA}$$\mid$ = 0.075. 
For the previous fixed mass and radius of the system this condition produced 
an initial temperature of the cloud T$_{cloud}$ $\approx$ 2200 K. The corresponding 
sound speed was $c~\approx~3.8$ km/sec. This is consistent with the typical measured 
``kinetic'' temperatures for such GMC complexes \cite{DHT2001}. 

With these parameters we have an initial central concentration of 
$n_0$ $\approx$ 10$^3$ cm$^{-3}$, and a free-fall time in the cloud 
center of $\tau_{ff}$ $\approx$ 1 Myr. The central Jeans radius is 
R$_{\rm J}$ $\approx$ 10 pc with the corresponding Jeans 
mass of M$_{\rm J}$ $\approx$ 10$^5$ M$_{\odot}$. Initially we give the 
whole system a rigid rotational velocity distribution with an angular velocity 
value $\Omega_{cloud}$ which we set to the unity of 
$\Omega_{0}$ = V$_{0}$/ R$_{cloud}$ where:

$$
{\rm V}_{0} \equiv \sqrt{ {\rm G}~{\rm M}_{cloud}/{\rm R}_{cloud} }.
$$

Using our parameters this velocity is equal to V$_{0}$ $\approx$ 21 km/sec.

The main rotational parameters ($\omega$ and $\beta$) for our two sets of 
models are listed in Table~\ref{tabl}.

\begin{table}[htbp]


\caption{\small The list of initial ``rotational'' parameters in our models}
\label{tabl}
\begin{center}
\begin{tabular}{llllllll}
\hline\noalign{\smallskip}
$\omega$ $\equiv$ $\Omega_{cloud}/\Omega_{0}$ & = & 0.05 & 0.10 & 0.15 & 0.20 & 0.25 & 0.30 \\
\noalign{\smallskip}
\hline
\noalign{\smallskip}
$\beta$ $\equiv$ {\rm E}$_{\rm KIN}$/E$^0_{\rm GRA}$ & $\approx$ & 6.2~10$^{-4}$ & 2.5~10$^{-3}$ & 5.6~10$^{-3}$ & 1.0~10$^{-2}$ & 1.5~10$^{-2}$ & 2.2~10$^{-2}$ \\
\noalign{\smallskip}
\hline
\end{tabular}
\end{center}


\end{table}

As we can see from Table~\ref{tabl} the initial ratio of the rotational (or kinetic) energy of the motion the the fragments ($\beta$), even in a last models with ``high'' rotational parameter, don't exceed more than a 
few \% from the gravitational bounding energy of the cloud. This is even less for the initial ratio of the 
systems thermal energy to the gravitational energy ($\alpha$ = 7.5~\%). In all models, usually after the first few Myr of evolution, these situations have changed. The ratio $\beta$ is rising to the approximate value of 0.5 or even more. Its mean is when the cloud starts the process of intensive isothermal fragmentation and the whole system of fragments becomes almost fully ``rotationally'' supported.




\bigskip
\noindent {\small {\bf RESULTS}}
\medskip

\noindent In Fig.~\ref{cluster} we show the time evolution of the total 
number of clusters N$_{\rm c}$ and the total mass fraction inside these 
clusters $\phi$ during the simulations. Starting from $\sim$ 5 Myr 
more than 80 \% of the total mass is already concentrated inside the fragments. 
At around 6 Myr already almost 95 \% of the total mass is inside the clusters.
Fig.~\ref{cluster}a shows the results for the ``slow'' rotating 
model with $\omega = 0.1$. In Fig.~\ref{cluster}b we show the evolution 
of the ``fast'' rotating model with $\omega = 0.3$.

In Fig.~\ref{icmf} we show four different snapshots of the integrated 
cluster distribution function (ICMF) for two selected moments of time with 
two different rotational parameters $\omega$. For practical numerical reasons 
we use the ICMF instead of the CMF (which is sometimes in the literature 
also called the Differential Cloud Mass Function DCMF). Here is a 
simple definition of the ICMF:

$$
{\rm ICMF} = \int_{0}^{M} d{\rm N}/d{\rm M}.
$$

Basically it shows how many clouds we have from zero mass to any 
fixed mass (M). Because the CMF is usually approximated with the power 
law: ${\rm M}^{-\gamma}$ in this case the ICMF we can be simply derived 
as $\sim {\rm M}^{-\gamma + 1}$.

The reason for using the ICMF instead of the CMF (DCMF) is that the 
averaging and slope definition is mathematically better due to the 
integrated CMF (which is a monotone function) and because the histograms 
in this case don't have any ``holes''. Of course when we compare 
our results with the observed (differential) CMF slope we need to subtract 
one from the ICMF slope to get the corresponding CMF slope.

In Fig.~\ref{icmf} we can see that in most cases the ICMF slope lies 
between -0.5 and -0.7 (the corresponding CMF is -1.5 and -1.7). The models 
with slow rotation always have a significantly lower value of the slope.

The ICMF slope time evolution for the set of our models with different 
rotation parameters are presented in Fig.~\ref{slope}. The models with 
initial ``slow'' rotational parameters give the ICMF average slope a level 
of -0.8 (which corresponds to the CMF slope -1.8). The ``fast'' 
rotating models give the ICMF slope a level of -0.4 (CMF slope -1.4). 

On average all models show very close values of the CMF slopes in comparison 
with the observed values of $\gamma$ in the different parts of our Galaxy. 

The slow rotation models systematically show the slope more close to the 
observed values in the outer part of our Galaxy \cite{KSRC1998, HCS2001, DHT2001} ($\gamma~\approx$~-1.8$\pm$0.03). In contrast, the fast model CMF slopes is 
more consistent with the observations from the central part of our 
Galaxy \cite{SRBY1987, SSS1985, MT2000} $\gamma~\approx$~-1.5$\pm$0.1.

Of course our simulations are time and also resolution limited, but even in 
this case we can derive a statement about the two significantly different 
types of ``population'' in the molecular cloud distributions. The key parameter 
which produces the different CMF slopes is the initial rotational 
parameter of the forming (and subsequently fragmenting) GMC.



\newpage


\bigskip
\noindent {\small {\bf CONCLUSIONS}}
\medskip

\noindent In this paper we present a study of the subsequent 
(runaway) fragmentation of the rotating isothermal GMC complex. 
Our own developed {\tt GRAPE} based Smoothed Particle 
Hydrodynamics (SPH) gas-dynamical model successfully reproduced 
the observed Cloud Mass-distribution Function (CMF) in our Galaxy. 
The steady state CMF is quickly established during the collapse 
approximately on a scale of a few free-fall time in the central 
parts of the modeled GMC. 

One of the key points in our model is that using our results we 
can naturally explain the source of possible differences 
between the observed slope on molecular clouds mass distribution 
function in the Galactic center and the outer regions of our Galaxy. 

The basic idea, is what if the GMC formed as a result of the 
galactic disk instability on the scale of the disk height ($\sim$100 pc). 
In such a case the initial angular momentum of the forming GMC 
can be defined by the Coriolis force during the formation inside the 
differentially rotating disk. Therefore the central GMC has a bigger $\beta$ 
and the external GMC has a smaller rotational parameter. 

According to our models this produces the different slopes of the 
resulting CMF during the runaway fragmentation process inside the 
system. The observed CMF gives to the central parts of Galaxy a 
slope well approximated with the value $\gamma~\approx$~-1.5$\pm$0.1 
\cite{SRBY1987, SSS1985, MT2000} and for the outer parts of the Galaxy 
the approximate value $\gamma~\approx$~-1.8$\pm$0.03 
\cite{KSRC1998, HCS2001, DHT2001}. 

Our results for the ``slow'' and ``fast'' rotating models give us exactly 
the same slopes with very good agreement with the recent 
observations. The ``slow'' models corresponds to the initially more 
slowly rotating GMC in the outer parts of the Galaxy. The ``fast'' 
rotating models corresponds to the GMC in the central part of the Galaxy. 
The central GMC can initially get more angular momentum from the differential 
rotation of the galactic disk during the process of GMC formation 
itself.

Our numerical investigation clearly shows that one of the key parameters, which 
determines the observed slope of the present day molecular CMF in different 
parts of our Galaxy, is the initial ratio of the rotational (turbulent) and 
gravitational energy inside the forming GMC.



\bigskip
\noindent {\small {\bf ACKNOWLEDGEMENTS}}
\medskip

\noindent P.B. wish to express his thanks for the 
support of his work to the German Science Foundation 
(DFG) under the grant {\bf SFB-439} (sub-project {\bf B5}). 
P.B. work was also supported by the following grants: 
{\bf NNG04GJ48G} from NASA, {\bf AST-0420920} from NSF 
and by {\bf HST-AR-09519.01-A} from STScI. He is very grateful 
for the hospitality of the Astronomisches Rechen-Institut 
(Heidelberg, Germany) where the part of these work has 
been done. The work of the authors was also supported by the 
Ukrainian State Fund of Fundamental Investigation 
under the project {\bf 02.07.00132}.

The calculation has been computed with the Mitaka 
Underground Vineyard (MUV) {\tt GRAPE6} system of the 
National Astronomical Observatory of Japan. The authors 
are want to express the special thanks for our 
colleagues Naohito Nakasato (Computational Astrophysics 
Group, RIKEN) for his constant help and support in the 
process of using the NAOJ {\tt GRAPE6} computational 
facilities. 

The authors are also very grateful to Dan Batcheldor 
(Astrophysics Group, RIT) for his constructive 
comments to the first variant of the paper.



\end{document}